\documentclass[3p,times]{elsarticle}

\usepackage{amssymb}
\usepackage{amsmath, bbm}
\usepackage[figuresright]{rotating}
\usepackage{capt-of}

\newcommand{\La}{{\Lambda}}
\newcommand{\Si}{{\Sigma}}

\newcommand{\be}{\begin{eqnarray}}
\newcommand{\ee}{\end{eqnarray}}

\newcommand{\LL}{{\Lambda\Lambda}}

\newcommand{\Kmp}{{K^-p}}

\newlength{\feynwidth} 
\newlength{\feynwidthbig} 

\usepackage{xcolor} 

\begin{document}

\begin{frontmatter}
  

\title{Coupled-channel effects in hadron-hadron correlation functions}
\author[J]{J. Haidenbauer}
\address[J]{Institute for Advanced Simulation, Forschungszentrum J{\"u}lich, D-52425 J{\"u}lich, Germany}
\begin{abstract}
Two-particle momentum correlation functions as measured in heavy ion collisions or in
high-energetic proton-proton collisions are studied. Special emphasis is put on systems like 
$\Lambda\Lambda$ or $K^-p$ where effects from the coupling to other channels could be relevant.
In both cases other channels open at relatively low momenta or are already open
at the reaction threshold. To have a solid basis, realistic coupled-channel interactions for 
$\Lambda\Lambda-\Xi N-\Lambda\Sigma-\Sigma\Sigma$ and $\pi\Lambda-\pi\Sigma-\bar KN$ 
are utilized in the actual calculations. It is found that the opening of the $\Xi N$ channel 
leaves a trace in the $\Lambda\Lambda$ correlation function that could be detectable in experiments. 
Should the proposed $H$-dibaryon be located close to or below the $\Xi N$ it will have a very 
pronounced effect. 
The presence of open channels in systems like $\Xi^- p$ or $K^-p$ does influence the 
correlation functions significantly at low momenta and will certainly complicate any 
dedicated analysis.  
\end{abstract}
\end{frontmatter}

\section{Introduction}

Traditionally, the determination of two-particle momentum correlations in relativistic heavy ion 
collisions has been viewed as a tool to explore properties like the size of the emitting source 
or the time dependence of the emission process \cite{KP1,LL1,KP2,Bauer:1992,LL2}. 
In recent times, however, another aspect has come more to the fore, namely the possibility to 
study and extract information on the interaction between the two particles  
\cite{Kisiel:2014,Shapoval:2015,Morita:2015,Ohnishi:2016,Mihaylov:2018}. 
After all, those correlations do not only depend on quantum statistical effects and on the
source but also reflect the final-state interaction (FSI) of the emitted particle pair. 
With that aim, measurements of the momentum correlations for baryon-baryon systems such as 
$p\Lambda$~\cite{Adams:2006,Anticic:2011,Adamczewski:2016,Acharya:2018},
$\bar{p}\La\, (p\bar{\La})$~\cite{Adams:2006}, and 
$\Lambda\Lambda$~\cite{Acharya:2018,STAR,Adam:2016} 
have been performed over the last few years, in heavy ion collisions but also in 
high-energetic proton-proton ($pp$) collisions, and analyzed theoretically 
\cite{Kisiel:2014,Shapoval:2015,Morita:2015,Ohnishi:2016}. 
See also Ref.~\cite{Cho:2017} for a recent review. 

Among the studied systems the $\Lambda\Lambda$ channel is certainly the most interesting 
one, not least due to the fact that the famous $H$-dibaryon \cite{Jaffe:1976} was predicted 
to be present here. Its quantum numbers coincide with that of the $\Lambda\Lambda$ state 
in the $^1S_0$ partial wave.
Given that direct scattering experiments are impossible and binding energies of 
observed double-$\Lambda$ hypernuclei like the so-called Nagara event 
${}_{\Lambda\Lambda}^{\;\;\;6}{\rm He}$~\cite{Takahashi:2001} 
can provide only rough constraints, FSI effects \cite{Gasparyan:2004} open an important and 
selective way to the strength of the $\Lambda\Lambda$ interaction \cite{Ohnishi:2016,Gasparyan:2012} 
and could have the potential to yield solid quantitative information. 
The same applies to baryon-baryon systems like $\Xi N$, $\Omega N$, or $\Omega\Omega$ 
where measurements of the corresponding correlation function are planned or already under 
way \cite{Shah:2017,Fabbietti:2018} 
and predictions can be found in the literature \cite{Mihaylov:2018,Morita:2016,Hatsuda:2017}. 
 
Various aspects concerning the practical application of the pertinent formalism to systems 
like $\Lambda\Lambda$ have been investigated thoroughly in recent times and are
well documented in published 
works~\cite{Kisiel:2014,Shapoval:2015,Morita:2015,Ohnishi:2016,Mihaylov:2018}. 
However, there is one specific feature which has not been dealt with in detail so far, 
namely possible effects from channel couplings. Specifically, in case of $\Lambda\Lambda$
the first inelastic channel, $\Xi N$, is only separated by about $25$~MeV, and is likely
to influence the results for the correlation function. Conversely, the open $\Lambda\Lambda$
channel should have an effect on the correlation function for $\Xi N$. 
A similar situation arises for the $K^-p$ system where 
measurements of the correlation function are likewise under way \cite{Lea:2018}. 
Also here there are open channels ($\pi\La$, $\pi\Si$) and, as a peculiarity, 
a small but relevant splitting between the $K^-p$ and $\bar K^0 n$ thresholds
due to the mass differences between $K^-$ and $\bar K^0$ \cite{PDG}. 
It should be said that the theoretical framework for dealing with multichannel systems 
has been already worked out in Ref.~\cite{LL2}. But, to the best of our knowledge, 
calculations for baryon-baryon and meson-baryon systems, where the channel coupling is
taken into account explicitly, have not been performed yet. 

In this paper we present an exemplary study of the influence of coupled-channel effects 
on the two-particle momentum correlation function. Systems like $pp$ or $K^+ p$ are 
ideal for testing and applying the femtoscopic tools developed in the past, because there is only a 
single $S$-wave state where strong correlations are expected to occur for low momenta 
and the thresholds of other channels are far away from the energy region of interest. 
However, already for $\La p$ the situation is different. There are two $S$-waves
(the $^1S_0$ and the $^3S_1$) owing to the fact that the two baryons can be in 
a spin singlet or triplet state, where correlations arise. 
Moreover, the $\La p$ system
couples to the $\Si N$ channel whose threshold lies only about $77$~MeV away 
and that coupling plays an important role in understanding the $\La N$ and $\Si N$ 
interaction within phenomenological potential models \cite{Hai:2005,Rijken:2010} 
but also in works based on modern frameworks like chiral effective field theory \cite{Hai:2013}. 
The significance of the coupling to $\Si N$ is clearly visible in experiments where, 
in various reactions, a pronounced cusp-like structure has been detected in $\La p$ 
observables at the opening of the $\Si N$ channel \cite{Machner:2013}.  
The conditions are more delicate for $\La \La$ where the threshold of 
the $\Xi N$ channel is separated only by around $25$~MeV, as already mentioned above. 
Indeed, the opening of the $\Xi N$ channel is well within the energy/momentum region where 
measurements of the $\La \La$ correlation function have been performed \cite{Acharya:2018,STAR}. 
The opposite scenario is realized in the $\Xi^- p$ system where the other channel ($\La \La$)
opens already below the reaction threshold.  Consequently, the pertinent scattering 
amplitudes no longer fulfill strict two-body unitarity constraints and a simple
effective range expansion that forms the basis in a convenient and widely
used approach for analyzing the correlation function \cite{LL1,Cho:2017} no longer works. 
Indeed, because of the mass splitting between $\Xi^0$ and $\Xi^-$ \cite{PDG}, the 
$\Xi^0 n$ channel opens as well below the $\Xi^- p$ threshold.
A similar situation occurs for the $K^- p$ system. Also here, some channels open
already well below the $K^- p$ threshold ($\pi\La$, $\pi\Si$) while another one, 
namely $\bar K^0 n$, has its threshold just about $5$~MeV above that of $K^- p$. 

In order to explore the effect of channel coupling on the correlation functions 
we utilize baryon-baryon and meson-baryon interactions where the relevant channels
are explicitly included. This allows us to compute the wave functions for all those
channels explicitly and to utilize them in the actual evaluation of the correlation 
function within the so-called Koonin-Pratt formulation \cite{KP1,Cho:2017}. 
To be specific, we employ the hyperon-nucleon (YN) interaction derived in Ref.~\cite{Hai:2013}
in the framework of SU(3) chiral effective field theory (EFT), where the coupling
between $\La N$ and $\Si N$ is taken into account. For the baryon-baryon interaction
with strangeness $S=-2$ we take likewise an interaction based on chiral EFT \cite{Hai:2016}. 
In this case the coupling of the $\La\La$, $\Xi N$, $\La\Si$, and $\Si\Si$ systems
has been taken into account. 
Finally, for the $\bar K N$ interaction we resort to a meson-exchange potential that has
been constructed by the J\"ulich group \cite{MG,Hai:2011}. It includes the coupling to
$\pi\La$ and $\pi\Si$ among others. Since that model, published in 1992, is not in line
with recently established information on the $K^- p$ scattering length \cite{Bazzi:2011}, 
we consider also a refitted version where those constraints are incorporated.
Furthermore, we employ one of the so-called chirally motivated $\bar K N$ potentials by 
Ciepl\'y and Smekal \cite{Cieply:2012}, which implement results from unitarized chiral 
perturbation theory.
All calculations are performed for physical masses so that subtle effects like the
mentioned $\Xi^0 n$  - $\Xi^- p$ and $K^- p$ - $\bar K^0 n$ splittings can be taken
into account appropriately. 

The paper is structured in the following way: 
In the subsequent section we summarize the employed formalism. Since that formalism
has been described in detail in various dedicated publications we will be very brief
here. 
In Sect. III numerical results for the correlation function are presented.
Thereby, the emphasis is on the $\La\La$ and $\Xi N$ interactions and on the
$K^-p$ system. However, results for $\Si^+\Si^+$ and $\La p$ will be also briefly
discussed.
The paper ends with some concluding remarks. 

\section{Formalism}

The formalism for calculating the two-particle correlation function from a two-body interaction
has been described in detail in various publications 
\cite{KP1,LL1,KP2,Bauer:1992,LL2,Morita:2015,Ohnishi:2016,Cho:2017}. Thus, in the following we 
will be very brief and provide only an overview of the employed formulae. Thereby, we follow very 
closely the presentation in the paper of Ohnishi et al.~\cite{Ohnishi:2016}. 

The two-particle momentum correlation function is given by 
\begin{align}
C(\bold{p_1},\bold{p_2})
=&\frac{
\int d^4x_1 d^4x_2
S_1(x_1,\bold{p}_1)
S_2(x_2,\bold{p}_2)
\left| \Psi^{(-)}(\bold{r},\bold{k}) \right|^2
}{
\int d^4x_1 d^4x_2
S_1(x_1,\bold{p}_1)
S_2(x_2,\bold{p}_2)
}
\label{Eq:KP}\\
\simeq&
\int d\bold{r}
S_{12}(\bold{r})
\left| \Psi^{(-)}(\bold{r},\bold{k}) \right|^2 \ .
\label{Eq:Corr}
\end{align}
Here the quantity $S_i(x_i,\bold{p}_i)~(i=1,2)$ is the single particle source function
of hadron $i$ with momentum $\bold{p}_i$, and 
$\bold{k}=(m_2\bold{p}_1-m_1\bold{p}_2)/(m_1+m_2)$. 
As already indicated by Eq.~\eqref{Eq:Corr}, we evaluate the quantity in question in the
center-of-mass (c.m.) frame where the wave function $\Psi^{(-)}$ is then a function
of the relative coordinate $\bold{r}$ and the c.m. momentum  
and $S_{12}(\bold{r})$ is the normalized pair source function that depends likewise
only on the relative coordinate. Furthermore, we consider only interactions in the $S$-wave,
though part of the formalism below will be written down in a general way. 

Assuming a static and spherical Gaussian source with radius $R$, 
$S(x,\bold{p})\propto \exp(-\bold{x}^2/2R^2)\delta(t-t_0)$,
a partial wave expansion can be performed straightforwardly and the correlation function 
can be written in a compact form \cite{Ohnishi:2016}.
In particular, for systems with two non-identical particles such as $\Lambda p$, $\Xi^-p$,
or $K^-p$ the correlation function amounts to 
\begin{align}
C(k)\simeq
1+\int_0^\infty 4\pi r^2\,dr\, S_{12}(\bold{r})
\left[
\left|\psi(k,r)\right|^2
-\left|j_0(kr)\right|^2
\right] \ ,
\label{Eq:cni}
\end{align}
where the properly normalized source function is given by $S_{12}(\bold{r})=\exp(-r^2/4R^2)/(2\sqrt{\pi}R)^3$  
and $j_l(kr)$ is the spherical Bessel function for $l=0$. $\psi(k,r)$ is the scattering wave
function that can be obtained by solving the Schr\"odinger equation for a given potential, but
also from the Lippmann-Schwinger (LS) equation as we will discuss below. 
Since $S$-wave baryon-baryon states can be formed with total spin $0$ or $1$ 
(i.e. can be in the partial waves $^1S_0$ and $^3S_1$, respectively) an averaging over the spin 
has to be performed in Eq.~(\ref{Eq:cni}). Thereby it is usually assumed that the weight is the 
same as for free scattering, namely $1/4$ and $3/4$, respectively. 
See, however, the discussion in Ref.~\cite{Bauer:1992}.

For systems with two identical particles like $\Lambda \Lambda$ the correlation function 
is given by 
\begin{align}
C(k)\simeq
1 - \frac12 \exp(-4k^2R^2)+\frac12 \int_0^\infty 4\pi r^2\, dr\,
S_{12}(\bold{r})
\left[
\left|\psi(k,r)\right|^2
-\left|j_0(kr)\right|^2
\right] \ .
\label{Eq:cid}
\end{align}
Here, besides a contribution due to the actual interaction another term arises from a quantum 
statistical effect which suppresses the correlation due to the anti-symmetrization of the wave 
function \cite{Ohnishi:2016}.

The wave functions to be inserted in Eqs.~(\ref{Eq:cni}) and (\ref{Eq:cid}) are normalized
asymptotically according to \cite{Ohnishi:2016} 
\begin{align}
\psi(k,r) \to
 \frac{e^{-i\delta}}{kr}\sin(kr+\delta)
=\frac{1}{2ikr}\left[e^{ikr}-e^{-2i\delta}e^{-ikr}\right]
\quad (r\to \infty)\ ,
\label{Eq:wf}
\end{align}
where $\delta = \delta(k)$ is the phase shift.
This differs from the standard definition by an overall phase $e^{-2i\delta}$,
see e.g. Ref.~\cite{Joachain}, which, however, drops out anyway in the actual 
calculation because the absolute square has to be taken. 

Simple and handy expressions can be derived when one assumes that the wave function is given 
everywhere by the asymptotic form. 
This is done in the Lednicky and Lyuboshitz (LL) model \cite{LL1,LL2} where one
arrives at

\begin{align}
\int_0^\infty 4\pi r^2 dr\,S_{12}(r) 
\left[
\left|\psi(k,r)\right|^2
-\left|j_0(kr)\right|^2
\right] 
\approx 
\frac{|f(k)|^2}{2R^2} F(r_0) 
+\frac{2\text{Re}f(k)}{\sqrt{\pi}R}\,F_1(x)
-\frac{\text{Im}f(k)}{R}\,F_2(x) \ .
\label{Eq:LL}
\end{align}

Here $f(k)$ is the scattering amplitude which is related to the $S$ matrix by $f(k)=(S-1)/2ik$,
and in practical applications is often replaced by the effective range expansion, i.e.
$f(k)\approx 1/(-1/a_0 + r_0k^2/2 - i k)$ with $a_0$ and $r_0$ being the scattering length and
the effective range, respectively, where for the former the ``baryon-baryon'' sign convention 
is adopted. Furthermore, $F_1(x)=\int_0^x dt\, e^{t^2-x^2}/x$ and $F_2(x)=(1-e^{-x^2})/x$, 
with $x=2kR$. 
The factor $F(r_0) = 1-r_0/(2\sqrt{\pi}R)$ is a correction that accounts for the deviation of
the true wave function from the asymptotic form \cite{Shapoval:2015,Ohnishi:2016}.

Now we connect with our own formalism and conventions and describe 
how $r$-space wave functions can be evaluated from reaction amplitudes that are calculated in 
momentum space by solving the LS equation, as it is the case for our interaction potentials
for $\Lambda N$ \cite{Hai:2013}, $\Lambda\Lambda$, and $\Xi N$ \cite{Hai:2016}, and for $\bar KN$ scattering \cite{MG,Hai:2011}. 
To begin with we rewrite the asymptotic form (\ref{Eq:wf}) in terms of Bessel and Hankel functions
\cite{Joachain}, for arbitrary angular momentum $l$  
\begin{eqnarray}
\nonumber
\tilde\psi(k,r) &\to&  \frac{1}{2}\left[h_l^{(2)} (kr) + e^{ 2i\delta} h_l^{(1)} (kr)\right] \\
         &\to&  j_l (kr) - i \rho(k) T_l(k) \, h_l^{(1)}(kr) \ ,
\label{Eq:wfa}
\end{eqnarray}
where the wave functions in Eqs.~(\ref{Eq:wf}) and (\ref{Eq:wfa})  
are related by $\psi(k,r) = e^{-2i\delta} \tilde\psi(k,r)$. 
The on-shell reaction amplitude $T_l(k)$ introduced in Eq.~(\ref{Eq:wfa}) is related to
the $S$-matrix via $S_l = \exp(2i\delta) = 1- 2i \ \rho(k) \ T_l$, where
$\rho(k) = k \ E_1(k)E_2(k)/(E_1(k)+E_2(k))$ with
$E_i(k) = \sqrt{m_i^2 + k^2}$ being the energies of the particles $1$ and $2$.
In the non-relativistic case this reduces to $\rho(k) = k \ \mu_{12}$ with
the reduced mass $\mu_{12} = m_1 m_2 / (m_1 + m_2)$. 
In order to compute the wave function away from the asymptotic region one needs the 
reaction amplitude $T_l$ half-off-shell and one has to exploit the relations
$|\psi\rangle = |\phi\rangle + G_0 V |\psi\rangle$ and $V|\psi\rangle = T |\phi\rangle$,
cf. Refs.~\cite{Joachain} or \cite{Haftel:1970}, where $|\phi\rangle $ stands for
the free wave and $G_0$ is the free two-body Green's function.
Explicitly this reads for the single-channel case and after a partial-wave 
expansion
\begin{equation}
\tilde\psi(k,r) = j_l(kr) + \frac{1}{\pi} \int j_l(qr)\, dq q^2 
\frac{1}{E-E_1(q)-E_2(q)+i\epsilon} T_l(q,k;E) \ ,
\label{Eq:wfb}
\end{equation}
where $E$ is the total energy, i.e. $E=E_1(k)+E_2(k)$. 
Obviously, this Fourier-Bessel transform can be performed for $T$ matrices that result from any 
type of interaction, also for the ones of non-local potentials that typically arise in 
applications of chiral effective field theory \cite{Hai:2013,Hai:2016}.  

The extension to coupled channels or (angular-momentum) coupled partial waves is
straight forward. First we note that the relation between the $S$ and $T$ matrices 
is now
\begin{equation}
S_{\beta\alpha} = \delta_{\beta\alpha} -{2i} \sqrt{\rho_\beta\,\rho_\alpha}\, T_{\beta\alpha} 
\label{Eq:S}
\end{equation}
where $\rho_\alpha$ and $\rho_\beta$ are the corresponding phase-space factors in the incoming
and outgoing channels and $S$ and $T$ are now matrices in the channel space.
The asymptotic form Eq.~(\ref{Eq:wfa}) goes over into \cite{Shaw:1962}
\begin{eqnarray}
\nonumber 
\tilde\psi_{\beta\alpha}(r) &\to& \sqrt{\frac{\rho_\beta}{\rho_\alpha}}
\left(\delta_{\beta\alpha} j_l (k_\alpha  r) \ - \ i h_l^{(1)}(k_\beta r) \sqrt{\rho_\beta\,\rho_\alpha}\, T_{\beta\alpha}\right) \\
&\to& \frac{1}{2} \sqrt{\frac{\rho_\beta}{\rho_\alpha}} 
\left[\delta_{\beta\alpha} h^{(2)}_l (k_\alpha r) + h^{(1)}_l(k_\beta r) 
\left(\delta_{\beta\alpha} \ - \ 2 i \sqrt{\rho_\beta\,\rho_\alpha} \, T_{\beta\alpha} \right) \right]
\label{Eq:wfcc}
\end{eqnarray}
where again the index $\alpha$ stands for the incoming channel and $\beta$ for the outgoing channel. 
The normalization used for the correlation functions in Ref.~\cite{Ohnishi:2016} can be 
recovered by multiplying the wave function in Eq.~(\ref{Eq:wfcc}) (the part within
the square brackets) with $S^\dagger$ from the right, exploiting that the $S$ matrix in 
Eq.~(\ref{Eq:S}) is unitary.   

For arbitrary $r$ the wave functions for the different channels are calculated from an 
equation analogous to Eq.~(\ref{Eq:wfb}),
\begin{equation}
\tilde\psi_{\beta\alpha}(r) = \delta_{\beta\alpha} j_l(k_\alpha r) + \frac{1}{\pi} \int j_l(qr)\, dq q^2 
\frac{1}{E-E^\beta_1(q)-E^\beta_2(q)+i\epsilon} T_{\beta\alpha;\,l}(q,k_\alpha;E) \ ,
\label{Eq:wfc}
\end{equation}
were $T_{\beta\alpha}$ is the half-off-shell transition amplitude. Note that the integral 
contains the propagator (Green's function) for the final state which does not have a singularity 
for channels that are closed, i.e. for $E < m^\beta_1 + m^\beta_2$. 

The calculation of correlation functions for multi-channel systems has been discussed thoroughly 
by Lednicky et al. in Ref.~\cite{LL2}. According to that work the wave functions in
Eqs.~(\ref{Eq:cni}) and (\ref{Eq:cid}) have to be substituted by those that describe the
scattering of the particles in question in the coupled-channel context, i.e.
\begin{eqnarray}
|\psi(k,r)|^2 \to \sum_\beta \omega_\beta |\tilde \psi_{\beta\alpha}(r)|^2
\label{Eq:www}
\end{eqnarray} 
where the sum $\beta$ runs over all two-body channels that couple to the state $\alpha$
and that can occur as intermediate states. The quantity $\omega_\beta$ is the corresponding 
weight. Indeed, in principle, the source function could be different as well for the different
channels~\cite{LL2} so that the sum should be outside of the intergral in Eq.~(\ref{Eq:cni}).  
However, in the present exploratory calculation we avoid to introduce additional parameters
and we assume that the source is the same for all channels and, moreover, 
we assume that the weights are all the same and equal to $1$. 

For simplicity reasons we ignore the Coulomb interaction in this exemplary work.
However, in principle, it is possible to include the Coulomb force in the momentum-space
calculation by the Vincent-Phatak method~\cite{VP}, and to obtain the wave functions 
following the steps described in detail in Appendix D of Ref.~\cite{Holzenkamp:1989}.

\begin{figure}
\begin{center}
\includegraphics[height=115mm,angle=-90]{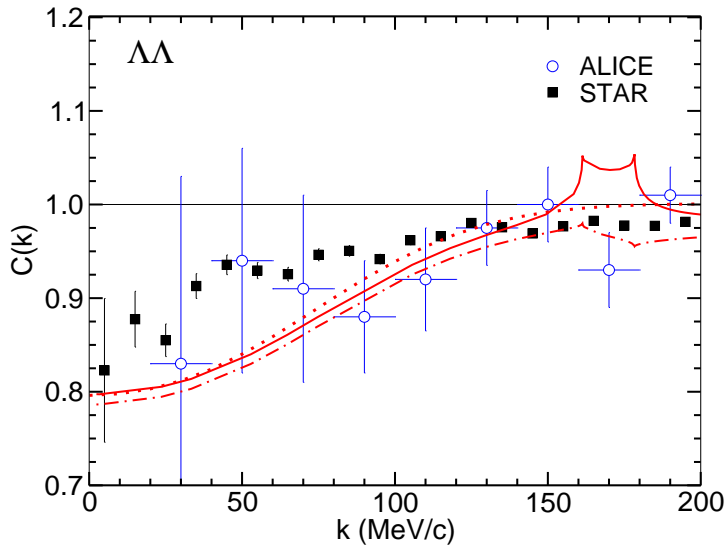}
\caption{Correlation function for $\LL$ evaluated for the NLO (600) interaction \cite{Hai:2016} 
based on the Koonin-Pratt formula and using physical masses.
The source radius is put to $R=1.2$~fm.
The dash-dotted line shows results when only the $\LL$ wave function is taken into
account while the solid line is the full calculation.  
The dotted line is based on the Lednicky-Lyuboshitz \cite{LL1,LL2} model formula,
Eq.~(\ref{Eq:LL}). 
Data are from the STAR~\cite{STAR} (squares) and ALICE \cite{Acharya:2018} 
(circles) Collaborations.
}
\label{fig:lala}
\end{center}
\end{figure}

\section{Results}

\subsection{$\La\La$, $\Xi N$ and $\Si\Si$ interactions}

To begin with we present results for baryon-baryon interactions with strangeness $S=-2$. 
That sector is rather rich as far as channel couplings are concerned. 
Even when only octet baryons are considered, as we do here, the $\La\La$ system can couple 
to $\Xi N$, whose thresholds lie only about $23$ MeV ($\Xi^0n$) and
$28$ MeV ($\Xi^-p$) higher 
(at a $\La\La$ c.m. momentum of $k\approx 161$ MeV/c and $178$ MeV/c, respectively),
and also to $\Si\Si$.
The $\Xi N$ channel itself can couple again to $\Si\Si$, but also to $\La\Si$. 
The $S=-2$ sector is also interesting for two other aspects. First, for the $\La\La$ channel 
actual data for the correlation function can be found in the literature~\cite{Acharya:2018,STAR}. 
Furthermore, the famous $H$-dibaryon \cite{Jaffe:1976} is predicted to be found in the 
$\La\La$ $^1S_0$ partial wave. 

Exemplary results for the $\La\La$ channel based on the chiral interaction in Ref.~\cite{Hai:2016} 
are shown in Fig.~\ref{fig:lala}, utilizing the next-to-leading order (NLO) potential with 
cutoff $\Lambda = 600$ MeV. There is only a single $S$-wave, the $^1S_0$, due to the 
Pauli principle. The predicted effective range parameters are 
$a_0 = -0.66$~fm, $r_0=5.05$~fm \cite{Hai:2016}. A summary of
the effective range parameters for all considered baryon-baryon channels is
provided in Table~\ref{ERE}. 
 
For orientation we include also available data from the STAR~\cite{STAR} and 
ALICE~\cite{Acharya:2018} collaborations in the figure, where the former are 
from a measurement of $\rm{Au} + \rm{Au}$ collisions at $200$~GeV and the latter
from $pp$ collisions at $7$~TeV. 
However, we refrain from performing an actual fit to those data. Indeed, this has been 
done already by others, cf. Refs.~\cite{Ohnishi:2016,Acharya:2018}, and it is re-assuring
to see that the effective range parameters deduced by Ohnishi et al.~\cite{Ohnishi:2016} 
are in line with those predicted by our NLO interaction. 
Note that the analysis of those data requires the introduction of additional 
parameters such as the so-called pair purity probability $\lambda$, normalization factors and 
possible corrections from residual correlations \cite{Kisiel:2014,Shapoval:2015,Ohnishi:2016},  
though, in principle, the value of $\lambda$ can be deduced from a detailed analysis of the 
experiment as outlined in the appendix of Ref.~\cite{Acharya:2018}. 
The calculation presented here does not involve any parameters,
once the source radius $R$ is fixed. For it we use $R=1.2$~fm throughout this work,
a value which is close or even identical to the one found in the analyses 
published in Refs.~\cite{Ohnishi:2016,Acharya:2018}. Anyway, it should not be concealed
here that the actual results for the correlation functions are quite
sensitive to the value of $R$ \cite{Ohnishi:2016,Mihaylov:2018}.

In Fig.~\ref{fig:lala} the prediction based on the $\La\La$ wave function
alone is indicated by the dash-dotted line while the solid line is the full results
that includes also the wave function for the $\La\La - \Xi N$ transitions. 
The calculation is performed with physical masses and one can clearly see the
opening of the $\Xi^0 n$ as well as of the $\Xi^- p$ channel in the
correlation function. The inclusion of the $\Xi N$ components clearly enhances
the signal around the corresponding thresholds. It will be interesting to see
whether future experiments with better resolution will be able to resolve 
the details here. The ALICE data \cite{Acharya:2018} included in Fig.~\ref{fig:lala}
and also earlier measurements of this collaboration \cite{Adam:2016} suggest that 
there could be indeed an anomaly at the $\Xi N$ threshold.  

\begin{table}[t]
\caption{Scattering lengths and effective ranges (in fm) for
the employed baryon-baryon interaction \cite{Hai:2016} 
for the various channels.
}
\label{ERE}
\vspace{0.2cm}
\centering
\renewcommand{\arraystretch}{1.5}
\begin{tabular}{|l||cc|cc|cc||cc|cc|}
\hline
         &\multicolumn{6}{|c||}{$^1S_0$} & \multicolumn{4}{|c|}{$^3S_1$} \\
\hline
         &\multicolumn{2}{|c|}{$I=0$} & \multicolumn{2}{|c|}{$I=1$}
         &\multicolumn{2}{|c||}{$I=2$} 
         &\multicolumn{2}{|c|}{$I=0$} & \multicolumn{2}{|c|}{$I=1$} \\
\hline
{channel} & $a$ & $r$ & $a$ & $r$ & $a$ & $r$ & $a$ & $r$ & $a$ & $r$ \\
\hline
$\Xi N $ & -10.89-\,i\,14.91 &  & 0.34 & -7.07 & & & -0.62 & 1.00 & 0.02 & 1797 \\
$\La \La $ & -0.66 & 5.05 & & & & & & & & \\
$\Si \Si $ & & & & & -1.83 & 6.05 & & & & \\
\hline
\end{tabular}
\renewcommand{\arraystretch}{1.0}
\end{table}

\begin{figure}[t]
\begin{center}
\includegraphics[height=115mm,angle=-90]{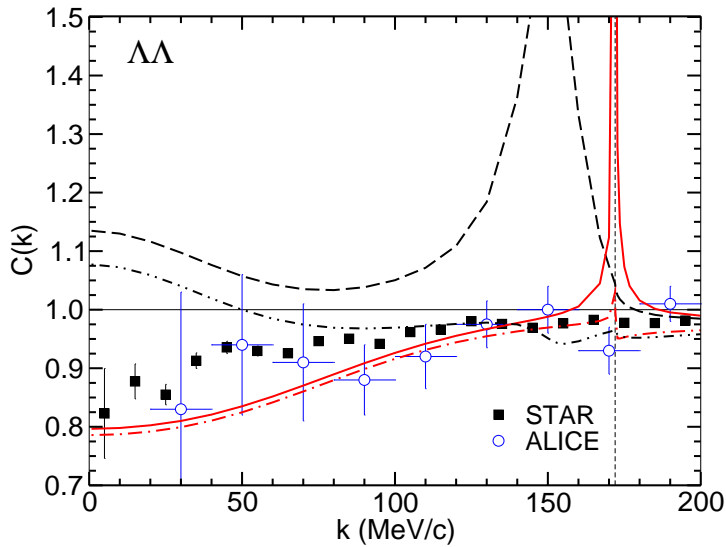}
\caption{Correlation function for $\LL$ evaluated for the NLO (600) interaction \cite{Hai:2016} 
based on the Koonin-Pratt formula and using isospin-averaged masses. 
The source radius is put to $R=1.2$~fm.
The dash-dotted line shows results when only the $\LL$ wave function is taken into
account while the solid line is the full calculation.  
The dash--double-dotted (dashed) line are corresponding results for a refitted $YY$ interaction 
that produces a $H$-dibaryon at a $\LL$ kinetic c.m. energy of $20$~MeV. 
The vertical line indicates the $\Xi N$ threshold. 
Data are from the STAR~\cite{STAR} (squares) and ALICE \cite{Acharya:2018} 
(circles) Collaborations.
}
\label{fig:lalai}
\end{center}
\end{figure}
 
An important observation with regard to the analysis of pertinent measurements is
that those $\Xi N$ components do not modify the $\La\La$ results for small momenta. 
This is not surprising 
because for energies below the $\Xi N$ threshold the corresponding wave functions 
drop exponentially so that their contributions should be suppressed.  
The same behavior emerges for the contribution of the $\Si\Si$ component
because in this case the corresponding threshold is much further away. In view of
that, and in line with the arguments given in the work of Lednicky et al.~\cite{LL2}, 
we omit the $\Si\Si$ wave function in the results shown in Fig.~\ref{fig:lala}. 
 
Another interesting aspect is that the approximation via the LL model (\ref{Eq:LL}),
utilizing the effective range parameterization for the evaluation of the scattering
amplitude $f(k)$, works very well for the $\La\La$ case -- despite the fact that the 
effective range is with $r_0=5.05$~fm significantly larger than the assumed source radius. 
In this context let us mention that we did also exploratory calculations for 
the strangeness $S=-1$ sector, and specifically for $\La p$ scattering
utilizing our corresponding NLO interactions \cite{Hai:2013}. Also there we 
found a very good agreement between the full calculations and the one based on 
Eq.~(\ref{Eq:LL}), for the $^1S_0$ as well as the $^3S_1$ partial waves. 
The opening of the $\Sigma N$ threshold has very little effect on the 
$\Lambda p$ correlation function -- which is not surprising in view of the
three times larger separation energy as compared to $\La\La-\Xi N$. 

In order to shed more light on the effect of the channel coupling we present here
also illustrative results for isospin-averaged masses. In this case there is only a single $\Xi N$
threshold and then a very pronounced cusp effect arises as can be seen in the
corresponding results of the EFT interaction for the $\La\La$ $^1S_0$ phase shift, 
cf. Fig.~6 (left side) in Ref.~\cite{Hai:2016}. This cusp is a remnant of the 
$H$-dibaryon which is predicted by the EFT potential as an inelastic 
virtual state \cite{Badalyan:1981}, and which lies very close to the $\Xi N$
threshold in the calculation for averaged masses
 -- see the pertinent discussion in \cite{Hai:2016}.
Obviously, as visible in Fig.~\ref{fig:lalai}, also in case of the correlation function the 
effect is much more drastic -- already when only the $\La\La$ wave function alone is used but 
even more so in the full calculation.

Recent lattice QCD calculations based on quark masses close to the physical
point \cite{Sasaki:2017,Sasaki:2018} but also extrapolations of older lattice results 
\cite{JMH1,JMH2,Inoue11a,Shanahan:2013,Yamaguchi:2016} suggest that a 
structure associated with the $H$-dibaryon could be present close to 
and somewhat below the $\Xi N$ threshold.
Because of that we re-adjusted slightly one of the low-energy constants 
of our chiral potential (the one corresponding to the SU(3) flavor singlet,
$C^1$, cf. Ref.~\cite{Hai:2016}) to turn the virtual state into a 
bound state, in order to explore its effect on the correlation function. We fixed 
the energy of the (unstable) bound state somewhat arbitrarily to be at about $5$~MeV 
below the $\Xi N$ threshold (i.e. $20$~MeV above the $\La\La$ threshold). 
The corresponding results are indicated by the dashed and dash--double-dotted
lines in Fig.~\ref{fig:lalai}. It is obvious that the presence of such a
bound state has an extremely strong effect on the correlation function and
one would really exclude the existence of such a state based on the
presently available data. Assuming that the bound state is 
located somewhat closer to the $\La\La$ threshold even increases the discrepancy 
between the predicted correlation function and the measurements. Indeed, even a
strong cusp effect as produced by the published NLO interaction \cite{Hai:2016}
(in the illustrative calculation with averaged masses) is practically ruled out by 
the presently available data. 

\begin{figure}
\begin{center}
\includegraphics[height=115mm,angle=-90]{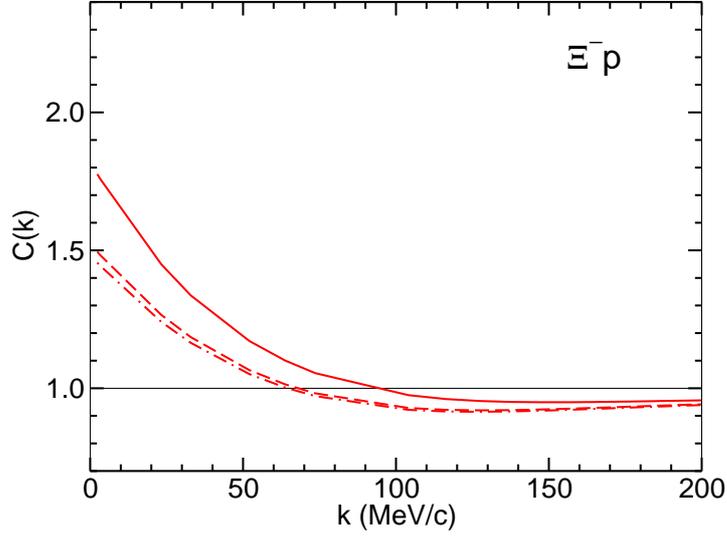}
\caption{Correlation function for $\Xi^- p$ evaluated for the NLO (600) interaction \cite{Hai:2016} 
based on the Koonin-Pratt formula. 
The source radius is put to $R=1.2$~fm.
The dash-dotted line shows results when only the $\Xi^-p$ wave function is taken into
account, the dashed line includes the $\Xi^-p-\La\La$ component, while the solid line 
includes also the $\Xi^-p-\Xi^0n$ component. 
}
\label{fig:xmp}
\end{center}
\end{figure}

Possible effects from an $H$-dibaryon located between the $\La\La$ and $\Xi N$
threshold have been also considered in Ref.~\cite{Morita:2015}. The effects
reported in that work are quite different from ours and much more modest. It 
should be said, however, that in \cite{Morita:2015} the $H$-dibaryon is added 
phenomenologically, namely in form of a Breit-Wigner distribution. In such a case, 
two-body unitarity of the $\La\La$ amplitude is no longer fulfilled. 
In our coupled-channel approach \cite{Hai:2016}, but also for the potential deduced
from LQCD simulations \cite{Sasaki:2018}, the resulting amplitudes fulfill two-body 
unitarity. Then the presence of such a (unstable) bound state has unavoidably drastic 
consequences for the $\La\La$ phase shifts and, in turn, for the wave functions 
and the correlation functions evaluated from them. 

\begin{figure}
\begin{center}
\includegraphics[height=115mm,angle=-90]{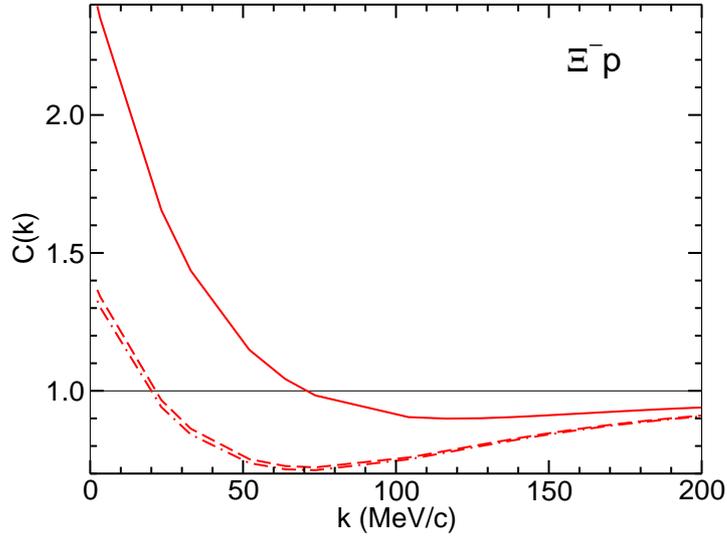}
\caption{Correlation function for $\Xi^- p$ evaluated for an alternative NLO (600) 
interaction discussed in Ref.~\cite{Hai:2016}, which produces a 
$\Xi N$ bound state in the isospin $I=0$ $^3S_1-{}^3D_1$ partial wave.
Same description of curves as in Fig.~\ref{fig:xmp}. 
}
\label{fig:xmpO}
\end{center}
\end{figure}

Results for the $\Xi^- p$ channel are shown in Fig.~\ref{fig:xmp}. 
There are two $S$-wave states, the $^1S_0$ and the $^3S_1$,
and we took the spin average in our calculation. Again we present 
predictions based on the $\Xi^- p$ wave function alone (dash-dotted curve)
and including the other wave functions, which in the present case are the ones
for $\Xi^- p - \Xi^0 n$ and $\Xi^- p - \La\La$. In accordance with 
the Pauli principle, the latter contributes only to the $^1S_0$ partial 
wave. Furthermore, it turned out that its contribution is rather small (dashed line). 
The difference between the dash-dotted and solid lines in Fig.~\ref{fig:xmp} is 
basically due to the $\Xi^- p - \Xi^0 n$ component alone.  
 
Obviously, and not unexpectedly, the contribution of the $\Xi^- p - \Xi^0 n$ component 
is sizable. Therefore, unlike the situation for $\La\La$ discussed above, 
a direct determination of the $\Xi^- p$ amplitude or scattering length from an empirical
$\Xi^- p$ correlation function is not really feasible. The empirical $\Xi^- p$ correlation 
function will alway involve contributions from both components, and, in principle, even with 
unknown weights, cf. Eq.~(\ref{Eq:www}). Actually, the basic drawback is 
well illustrated by the consideration of the reactions $\pi^+\pi^- \to \pi^+\pi^-$ and 
$\pi^+\pi^- \to \pi^0\pi^0$ within the LL model in Ref.~\cite{LL2}, where one
can see that different combinations of the elementary $\pi\pi$ amplitudes (assumed
in that work to have well-defined isospin) determine the scattering amplitude and 
the correlation functions, and they cannot be disentangled.
The very same applies to the $\Xi N$ system. 

The contributions of channels with higher thresholds ($\Lambda\Sigma$
which opens at $k\approx 230$ MeV/c, and $\Si\Si$) are not included in the results 
shown in Fig.~\ref{fig:xmp}. Their effect is fairly small, cf. the arguments given 
above. Note that, besides the hadronic interaction, in the $\Xi^- p$ 
channel there will be effects from the attractive Coulomb interaction.
As already said above we ignore the Coulomb force in this exploratory 
study. We expect that the Coulomb interaction will modify the correlation
function for momenta below $k\approx 50$ MeV/c, based on results presented
in Ref.~\cite{Bauer:1992}, and it should lead to a significant enhancement 
for decreasing $k$ values. 

The NLO potential used for the calculation shown in Fig.~\ref{fig:xmp} was constructed
in order to meet all available experimental constraints on the $\Xi N$ cross 
sections \cite{Hai:2016}. As a result the interaction is, in general, fairly weak
and specifically only weakly attractive in both ($I=0$ and $1$) $^3S_1$ partial waves. 
Indeed, its properties are very similar to the $\Xi N$ results reported from recent 
lattice QCD simulations close to the physical point \cite{Sasaki:2017,Sasaki:2018}. 
In contrast, potentials derived within the meson-exchange framework are usually
more strongly attractive and often lead to $\Xi N$ bound states,
in one or even more of the $S$-wave states \cite{Rijken:2010}. 
A similar tendency emerges within chiral EFT if one implements SU(3) flavor 
symmetry strictly, see the discussions in Refs.~\cite{Hai:2016,Hai:2014}. 
It should be said that such more strongly attractive $\Xi N$ interactions seem to be 
favored by studies of the observed spectrum of the $(K^-,K^+)$ reaction on a $^{12}$C 
target \cite{Khaustov}, which point to an attractive $\Xi$ single-particle potential of 
$U_\Xi \approx -14$~MeV \cite{Gal:2016}. Further support comes from 
evidence for the existence of deeply bound state in systems like 
$\Xi^-$-$^{\,14}$N \cite{Nakazawa:2015} or $\Xi^-$-$^{\,11}$B \cite{Nagae:2017},
reported within the last few years. 

In view of that controversial situation we present here also correlation functions
for an alternative NLO interaction, discussed in Ref.~\cite{Hai:2016} but finally dismissed
because it does not meet the empirical constraints on the $\Xi^-p \to \Xi^-p$ and
$\Xi^-p \to \Xi^0n$ cross sections. The corresponding results are displayed in 
Fig.~\ref{fig:xmpO}. This potential produces a $\Xi N$ bound state in the $I=0$ $^3S_1-{}^3D_1$
partial wave with a binding energy comparable to that of the deuteron~\cite{Hai:2016}.
There is a clear difference between the correlation functions predicted by an only
weakly attractive $\Xi N$ interaction and those that follow from a strongly
attractive force, cf. Figs.~\ref{fig:xmp} and \ref{fig:xmpO}. Thus, measurements of
that correlation function \cite{Fabbietti:2018} could allow one to distinguish between 
these two scenarios. 

\begin{figure}
\begin{center}
\includegraphics[height=115mm,angle=-90]{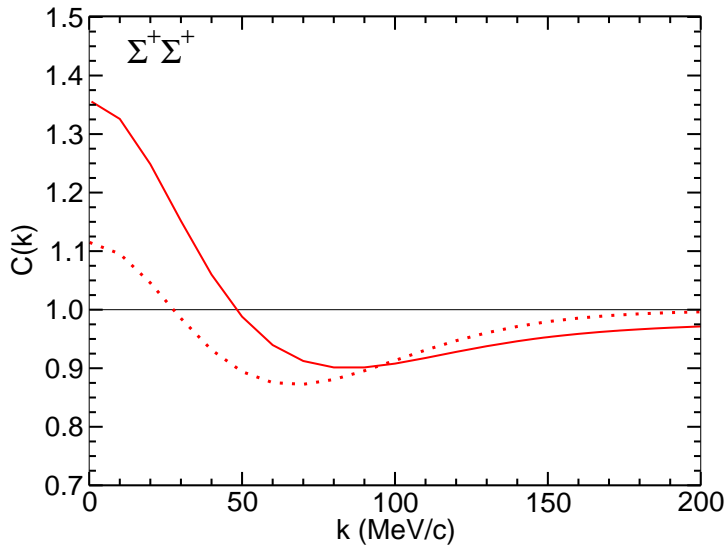}
\caption{Correlation function for $\Si^+\Si^+$ evaluated for the NLO (600) interaction \cite{Hai:2016} 
based on the Koonin-Pratt formula (solid line) and on the LL formula (dotted line),
Eq.~(\ref{Eq:LL}). The source radius is put to $R=1.2$~fm.
}
\label{fig:sisi}
\end{center}
\end{figure}


For completeness we provide also predictions for the $\Si^+\Si^+$ 
interaction, see Fig.~\ref{fig:sisi}. (Those for $\Si^-\Si^-$ are practically
identical under the assumption of charge symmetry.) This channel is 
interesting because, like $\La\La$, there is only a single $S$-wave, namely 
again the $^1S_0$. Moreover, for that system with total charge $Q=2$ there 
is no channel coupling so that the application of the formalism for 
femtoscopy \cite{Ohnishi:2016} is straight forward. 
Still our results indicate that it could be problematic to use the LL model.
Obviously, for fairly large effective range parameters as predicted by our 
interaction \cite{Hai:2016}, cf. Table~\ref{ERE}, a reliance on the effective 
range expansion in the application of Eq.~(\ref{Eq:LL}) is problematic.
Finally, note that the repulsive Coulomb interaction will suppress the
correlation function for small $k$ values, see, e.g., the situation for the $pp$ 
correlations in Ref.~\cite{Acharya:2018}. 

Of course, experimentally the measurement of $\Si\Si$ correlation functions 
is rather challenging since $\Si^+$ and $\Si^-$ are difficult to identify in
high-energy collisions given that their decay involves one neutral particle.
Nonetheless, experimental information on the ($\Si^+\Si^+$ or $\Si^-\Si^-$) 
scattering lengths would certainly provide a useful test for the SU(3) flavor 
symmetry and its possible breaking \cite{Hai:2016}. 
The $\Si^-$ and its interaction plays also an important role in the 
ongoing discussions on the properties of neutron stars, in the context
of the so-called hyperon puzzle 
\cite{Chatterjee:2015,Oertel:2016,Tolos:2016}.


\subsection{$\bar K N$ interaction}
Let us now come to the $K^- p$ system. There are already two channels
open at the $\bar K N$ threshold, namely $\pi\La$ and $\pi\Si$. Moreover, 
there is a relatively large mass splitting between the $K^-$ and $K^0$ so that
the $\bar K^0 n$ channels opens at about $5$~MeV above the $K^- p$ threshold,
or in terms of the $K^- p$ c.m. momentum at $k\approx 58$ MeV/c. Naturally,
this mass splitting induces a sizable isospin breaking in the 
threshold region. Further channels that couple are $\eta\La$, $\eta\Si$, etc.,
but their thresholds are at significantly higher energies and we do not
consider their effect explicitly in the present study. 

\begin{figure}
\begin{center}
\includegraphics[height=115mm,angle=-90]{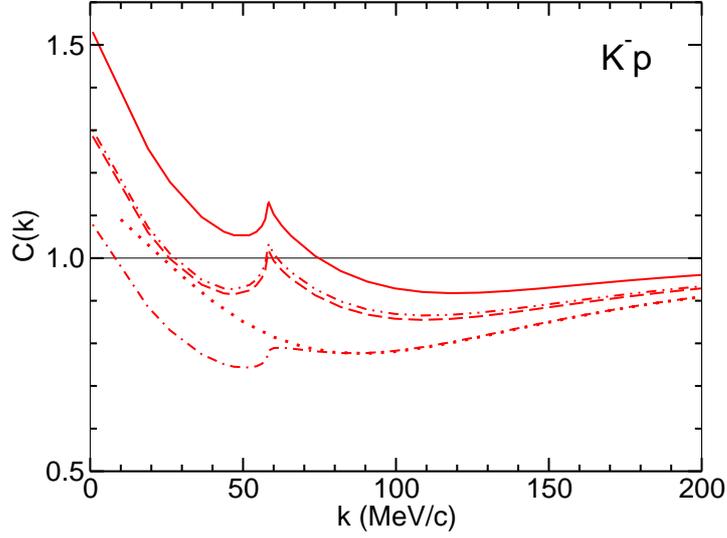}
\caption{Correlation function for $K^-p$ evaluated for  the original J\"ulich $\bar KN$ model \cite{MG} 
based on the Koonin-Pratt formula. 
The source radius is put to $R=1.2$~fm.
The dotted line is a calculation based on the $K^-p$ 
wave function alone and with isospin averaged masses, while the dash-dotted line is 
with physical masses. 
The dashed, dash--double-dotted and solid curves show results where 
contributions from the $K^-p - \bar K^0 n$, $K^-p - \pi^0 \Lambda$ and 
$K^-p - \pi \Sigma$ components are added consecutively. 
}
\label{fig:kmp0}
\end{center}
\end{figure}

We exemplify the influence of the various channels on the $K^- p$ correlation
function for wave functions generated from the J\"ulich meson-exchange 
potential \cite{MG,Hai:2011}. 
Pertinent results are summarized in Fig.~\ref{fig:kmp0}.
First, we performed a calculation with isospin-averaged masses (dotted line) where
then the $K^- p$ and $\bar K^0 n$ thresholds coincide and the wave function is 
simply a linear combination of the corresponding isospin $I=0$ and $I=1$ 
wave functions \cite{Ohnishi:2016}. 
The analog result but for physical masses is indicated by the dash-dotted line. 
In both cases only the $K^- p$ wave function is used. Obviously there is a drastic 
effect for momenta below the $\bar K^0 n$ threshold whereas at higher momenta the 
difference is very small. 
Adding the $\bar K^0 n$ component of the wave
function yields the dashed curve i.e. leads to a sizable shift upwards of the
correlation function and to a much more pronounced structure (cusp) at the $\bar K^0 n$ 
threshold. The effect of the $\pi^0\La$ component turns out to be basically negligible 
(dash--doubled-dotted line). However, adding finally $\pi\Si$ (which consists of
three components, namely $\pi^-\Si^+$, $\pi^0\Si^0$, and $\pi^+\Si^-$) leads
again to a strong modification of the $K^- p$ correlation function. 

\begin{figure}
\begin{center}
\includegraphics[height=115mm,angle=-90]{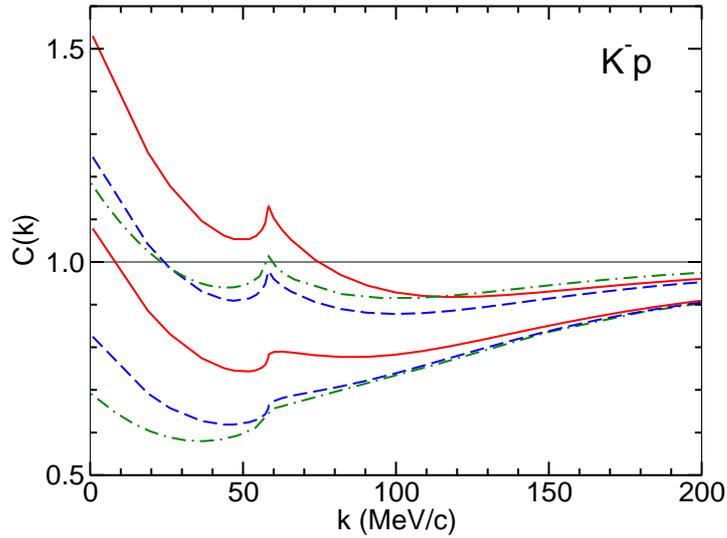}
\caption{Correlation function for $K^-p$ based on the $K^-p$ wave function alone 
(lower lines) and with inclusion of all channels (upper lines). 
Shown are results for the original J\"ulich $\bar KN$ model~\cite{MG} (red solid lines), 
a refitted J\"ulich $\bar KN$ model (see text) (blue dashed lines), 
and for the chirally motivated $\bar KN$ potential NLO30 \cite{Cieply:2012} (green dash-dotted lines). 
The source radius is put to $R=1.2$~fm.
}
\label{fig:kmp}
\end{center}
\end{figure}

The experimental determination of the level shift and width of kaonic 
hydrogen by the Siddharta Collaboration in 2011 \cite{Bazzi:2011} has put very 
tight constraints on the $K^-p$ scattering length.  Pertinent studies based on 
unitarized chiral perturbation theory \cite{Ikeda:2012,Mai:2013} together with 
an improved Deser formula \cite{Meissner:2004} quantify the latter now to 
$a_{\Kmp} = (-0.68\pm 0.18 + i 0.90\pm 0.13)$~fm \cite{Mai:2013}.
The J\"ulich model on the other hand, being from 1992, predicts 
$a_{\Kmp} = (-0.36+i 1.15)$~fm \cite{Hai:2011}.
Because of that we consider here also an up-to-date interaction where the result
for the $K^-p$ scattering length is in line with the value given above, 
namely the chirally motivated $\bar KN$ potential NLO30 by Ciepl\'y 
and Smejkal \cite{Cieply:2012}. It yields $a_{\Kmp} = (-0.75+i0.89)$~fm. 
The wave functions for this potential, which is given in separable form in
momentum space, can be calculated again via Eq.~(\ref{Eq:wfc}). 
Furthermore, we performed a rough refit of the J\"ulich $\bar KN$ model with the aim
to bring the $\Kmp$ results more in line with the value extracted from the level 
shifts. In this case a value of $a_{\Kmp} = (-0.70+i 1.13)$~fm has been achieved. 
Results for these two interactions are presented in Fig.~\ref{fig:kmp}, in
comparison to the original J\"ulich model. 
(Note that the scattering lengths given above conform now to the standard 
meson-baryon sign convention!)

The difference in the properties (predicted $K^-p$ scattering length) of the 
original J\"ulich model \cite{MG} and the refit are clearly reflected in the 
correlation functions. In particular, for the interaction with a realistic value
for $a_{\Kmp}$ the predicted $C(k)$ is significantly smaller at low $k$ values,
in the calculation where only the $K^-p$ wave function is used 
(cf. the lower dashed and solid lines in Fig.~\ref{fig:kmp}) 
as well as for the full calculation (upper dashed and solid lines). 
The result based on the chirally motivated NLO30 potential including only 
the contribution from the $K^-p$ wave function agrees remarkably well with that 
of the refitted J\"ulich interaction for momenta above the $\bar K^0 n$ threshold, 
cf. the lower dash-dotted and dashed lines in Fig.~\ref{fig:kmp}. 
The differences at very low momenta could reflect the noticeable variation in 
the corresponding scattering lengths, cf. above. 
With regard to the full calculation (upper solid, dashed and dash-dotted lines) 
the same trend is visible. Specifically, again the predictions for the
refitted J\"ulich interaction and the NLO30 potential lie fairly close together.

We believe that the large contribution from the $\pi\Si$ component is due to the
presence of the $\Lambda$(1405) resonance in the isospin $I=0$ channel.  
In potentials that incorporate chiral SU(3) dynamics of QCD like NLO30
but also in meson-exchange interactions, the $\Lambda$(1405) is generated 
dynamically by the strong attractive forces between the antikaon
and the nucleon. 
Actually, as a characteristic feature of chiral approaches even two 
poles are predicted in the region below the $\bar K N$ 
threshold \cite{Oller:2000,Cieply:2016,Hyodo:2012}.
The pole commonly identified with the $\Lambda$(1405) is located close to the 
$\bar K N$ threshold and leads to an enhancement not only in the $\bar K N$ 
amplitude itself but also in the $\bar K N-\pi\Sigma$ transition. This
feature is reflected in the resulting correlation function shown in 
Figs.~\ref{fig:kmp0} and \ref{fig:kmp}. Thus, via the correlation function it might 
be possible to have access to information complementary to that of $K^-p$ elastic scattering, 
as argued in Ref.~\cite{Ohnishi:2016}. Still, in practice it will be a challenge to
disentangle the impact that comes from the $\bar K N-\pi\Sigma$ coupling from 
the one that is caused by the transition $K^- p-\bar K^0 n$, where the latter
is moreover distorted by isospin-breaking effects. 
Clearly, both of those channel couplings affect the correlation function for small momenta. 

Note that wave functions from additional channels in the potentials that open at
higher energies
($\bar K \Delta$, $\bar K^* N$, and $\bar K^* \Delta$ in case of the J\"ulich potential \cite{MG}  
and $\eta \La$, $\eta \Si$, and $K\Xi$ in case of NLO30 \cite{Cieply:2012})
have been neglected. 

Finally, let us mention that predictions for the $K^- p$ correlation function have
been also presented in Ref.~\cite{Ohnishi:2016}, utilizing the $\bar K N$ potential
from Ref.~\cite{Miyahara:2016}. Those results look very different from the ones 
based on the J\"ulich and the NLO30 potentials shown here. A possible reason for this
could be that a much larger source radius was used in that study, namely $R=3$~fm. 
Newer (though still preliminary) results for this potential, now for practically the
same source radius, presented at the recent HYP2018 conference \cite{Lea:2018}, suggest 
that the results are more or less comparable, at least on a qualitative level.  

 \section{Conclusions}

In this work we have presented an exemplary study of two-particle momentum correlation functions 
as measured in heavy ion collisions or in high-energetic proton-proton collisions. 
Thereby, special emphasis has been put on systems like $\Lambda\Lambda$ or $K^-p$ where 
effects from the coupling to other channels could be of relevance.
Indeed, in both cases other channels open already at relatively low momenta or are already open
at the reaction threshold. To have a solid basis, realistic coupled-channel interactions for 
$\Lambda\Lambda-\Xi N-\Lambda\Sigma-\Sigma\Sigma$ and $\pi\Lambda-\pi\Sigma-\bar KN$ 
have been utilized. 
The pertinent wave functions for all relevant channels have been calculated explicitly and 
then employed in the actual evaluation of the correlation function within 
the Koonin-Pratt formulation \cite{KP1,Cho:2017}.  
 
Our study indicates that the opening of the $\Xi N$ channel should leave a trace in the 
$\Lambda\Lambda$ correlation function. Thus, experiments with improved statistics could
be able to resolve it and to shed light on the behavior of the $\Lambda\Lambda$ amplitude 
for momenta around the $\Xi N$ threshold. Specifically, 
if the proposed $H$-dibaryon is located close to or below the $\Xi N$ threshold, as 
indicated by the latest lattice QCD simulations near the physical point \cite{Sasaki:2018} 
and by extrapolations of older lattice results \cite{JMH1,JMH2,Inoue11a,Shanahan:2013,Yamaguchi:2016},
there should be a pronounced effect in the $\Lambda\Lambda$ correlation function. It should be
said, however, that presently available data \cite{Acharya:2018,STAR,Adam:2016} do not 
support the existence of a (bound) $H$-dibaryon in that energy region
but provide indications for a virtual state close to the $\Xi N$ threshold. 

With regard to the $\Lambda\Lambda$ correlation function at low center-of-mass momenta
we observe that its behavior is determined practically by the $\Lambda\Lambda$  
component of the wave function alone. The contribution from the $\Lambda\Lambda - \Xi N$
component is strongly suppressed. This facilitates the extraction of the $\Lambda\Lambda$
amplitude (effective range parameters) from pertinent measurements, as has been already
demonstrated in the literature \cite{Ohnishi:2016}. 
It is interesting to see that even the Lednicky-Lyuboshitz model \cite{LL1,LL2}
which utilizes only the asymptotic form of the wave function works very well in this
case. 
These findings are promising for application of the same formalism to other two-body 
interactions were similar kinematical conditions are realized like for $\Omega\Omega$ 
or systems with charmed baryons ($\Lambda_c p$, $\Lambda_c\Lambda_c$, ...) once
data become available. 
 
As far as systems like $\Xi^- p$ or $K^-p$ are concerned, our study reveals that the 
presence of open channels influences the correlation functions significantly for low momenta. 
This will certainly complicate any dedicated analysis. For example, it will be difficult
to draw quantitative and detailed conclusions on the $\Xi^- p$ amplitude itself
from pertinent experiments. However, issues like whether the $\Xi N$ interaction is 
only weakly \cite{Hai:2016,Sasaki:2018} or more 
strongly~\cite{Khaustov,Nakazawa:2015,Nagae:2017} attractive and, in particular, 
whether there are even bound states \cite{Rijken:2010} could be still resolved 
by analyzing empirical $\Xi^- p$ correlation functions. 
In case of the $K^-p$ system the measured level shift and width of kaonic hydrogen 
provide strong constraints on the $K^-p \to K^-p$ amplitude (wave function) 
and that knowledge can be utilized in the analysis of measured correlation functions.
However, since the coupling $\bar K N-\pi\Sigma$ as well as the coupling $K^- p-\bar K^0 n$ 
strongly influence the actual correlations for small momenta it will be a challenge to 
disentangle these effects in practice. 

\section*{Acknowledgements}
The author acknowledges stimulating discussions with Laura Fabbietti, Tetsuo Hyodo, 
Valentina Mantovani Sarti, Dimitar Mihaylov, and Ramona Lea. 
He also acknowledges communication with Ales Ciepl\'y concerning the chirally
motivated $\bar KN$ interaction. 
This work is supported in part by the DFG and the NSFC through
funds provided to the Sino-German CRC 110 ``Symmetries and
the Emergence of Structure in QCD'' (DFG grant no. TRR~110). 


\end{document}